\documentclass[11pt]{article}

\usepackage[labelfont=bf]{caption} % Bold figure number
\usepackage[font={small}]{caption}   %% FIGURE CAPTION FONT MODIFICATION

\usepackage{graphicx}% Include figure files
\usepackage{dcolumn}% Align table columns on decimal point
\usepackage{bm}% bold math
\usepackage{hyperref}
\usepackage[utf8]{inputenc}
\usepackage{lipsum}  
\usepackage{chngpage}
\usepackage{amsmath}
\usepackage{authblk}
\usepackage{amsfonts}
\usepackage{soul}
\usepackage{color}%
\usepackage{placeins}
\usepackage{comment}
\usepackage[numbers, sort&compress]{natbib}  %%% For compressing the reference numbering
\usepackage{lineno}
\begin{document}

\title{A fractional corner anomaly reveals higher-order topology}

\author[1]{Christopher W Peterson}
\author[2]{Tianhe Li}
\author[3]{Wladimir A. Benalcazar}
\author[2]{Taylor L. Hughes}
\author[4]{Gaurav Bahl}
\affil[1]{\footnotesize{Department of Electrical and Computer Engineering, University of Illinois at Urbana-Champaign, Urbana, IL, USA}}
\affil[2]{\footnotesize{Department of Physics and Institute for Condensed Matter Theory, University of Illinois at Urbana-Champaign, Urbana, IL, USA}}
\affil[3]{\footnotesize{Department of Physics, The Pennsylvania State University, University Park, PA, USA}}
\affil[4]{\footnotesize{Department of Mechanical Science and Engineering, University of Illinois at Urbana-Champaign, Urbana, IL, USA}}

\date{\today}

\maketitle

\begin{abstract}
Spectral measurements of boundary localized in-gap modes are commonly used to identify topological insulators via the bulk-boundary correspondence. This can be extended to high-order topological insulators for which the most striking feature is in-gap modes at boundaries of higher co-dimension, e.g. the corners of a 2D material.
Unfortunately, this spectroscopic approach is not always viable since the energies of the topological modes are not protected and they can often overlap the bulk bands, leading to potential misidentification.
Since the topology of a material is a collective product of all its eigenmodes, any conclusive indicator of topology must instead be a feature of its bulk band structure, and should not rely on specific eigen-energies.
For many topological crystalline insulators the key topological feature is fractional charge density arising from the filled bulk bands, but measurements of charge distributions have not been accessible to date.
In this work, we experimentally measure boundary-localized fractional charge density of two distinct 2D rotationally-symmetric metamaterials, finding 1/4 and 1/3 fractionalization. We then introduce a new topological indicator based on collective phenomenology that allows unambiguous identification of higher-order topology, even in the absence of in-gap states. Finally, we demonstrate the higher-order bulk-boundary correspondence associated with this fractional feature by using boundary deformations to spectrally isolate localized corner modes where they were previously unobservable.
\end{abstract}

\section{Introduction}
Topological insulators (TIs) are materials with a gapped band structure characterized by quantized quantities, called topological invariants, that are invariant under deformations that preserve both the bulk bandgap and any protective symmetries \cite{qi2011topological,chiu2016classification,schnyder2008classification}.
At a boundary between two materials that have different strong topological invariants, i.e., where a topological invariant changes in space, the bandgap closes and robust boundary-localized gapless modes appear.
Detection of these robust gapless boundary modes is therefore one of the most striking signatures of topological materials.
In this paper, we focus on two-dimensional TIs in class AI (spin-less and time-reversal symmetric) \cite{KitaevPeriodicTable2009,ryu2010topological,altland1997nonstandard}. 
In this class and dimension, no non-trivial strong topological invariants exist (i.e., those protected by particle-hole, chiral, and/or time-reversal symmetry, such as the $\mathbb{Z}_2$ invariant for a quantum spin Hall insulator in class AII), but invariants can be defined if additional spatial symmetries are present. 
Materials with invariants protected by spatial symmetries are known as topological crystalline insulators (TCI) \cite{TCIsreview,TeoSurface2008,FuTopological2011,FuTopological2012,SlagerSpace2013,ShiozakiTopology2014,FangNew2015,WatanabeTopological2017}.
Specifically, we are interested in a recently discovered class of TCIs whose members have gapped boundaries of codimension one, but host gapless modes at boundaries with codimension greater than one, i.e., at a boundary of a boundary.
Examples of these systems include 2D TCIs that support 0D corner modes \cite{QuadTheory,LangbehnReflection2017,song2017}, or 3D TCIs supporting 1D hinge or 0D corner modes \cite{benalcazar2017electric,HOTI,Bismuth,KhalafHigher2018}. 
Since these insulators manifest robust gapless modes at boundaries with higher codimension, they have been termed higher-order topological insulators (HOTIs).
One definition of an ($n$)-th order HOTI is a $d$-dimensional insulator that is gapless at ($d-n$)-dimensional boundaries, but gapped for all boundaries between ($d-n+1$) and ($d-1$)-dimensions.
We note that spatial symmetries are essential for these HOTIs, since they prevent bulk and surface deformations that hybridize and gap out the set of higher-order gapless states.
In the short time since they were first predicted, only a few naturally occurring HOTIs have been identified \cite{HOTI,Bismuth}.
Instead, much of the experimental study of HOTIs (primarily $d$-th order TCIs in $d$ dimensions) has been performed in engineered metamaterials such as networks of coupled resonators \cite{Peterson2018,ImhofTopolectrical2018,AcousticKagome1, AcousticKagome2,Mittal2019,Acoustic3rd1, Acoustic3rd2}, waveguide arrays \cite{Noh2018,Hassan2019}, and photonic or sonic crystals \cite{HuberQuadrupoleNature2018,SonicCrystalHOTI,lin2019nonsymmorphic,Xie2019,Chen2019}.
So far, the clearest indicator of higher-order topology in such systems has been the spectroscopic measurement of robust localized corner modes with energies inside the bulk bandgap of 2D \cite{Peterson2018,ImhofTopolectrical2018,AcousticKagome1, AcousticKagome2,Mittal2019,Noh2018,Hassan2019,HuberQuadrupoleNature2018,SonicCrystalHOTI,lin2019nonsymmorphic,Xie2019,Chen2019} and 3D \cite{Acoustic3rd1, Acoustic3rd2} HOTIs.
However, there exists a fundamental problem with using localized in-gap boundary modes to identify higher-order topology (or, in general, topology protected by spatial symmetries).
Spatial symmetries essentially divide a material into symmetric sectors, and require that localized modes in each sector are identical. Hence these symmetries protect the degeneracy of boundary-localized modes, but do not restrict their energy \cite{CornerCharge}.
Additional local symmetries (e.g., chiral symmetry or particle-hole symmetry) can pin the boundary modes to zero energy (mid-gap) \cite{CornerCharge,AhnFailure2019}, but these symmetries are not actually necessary to protect the higher-order topology, and many lattice models do not support their implementation at all.
This implies that the energy of localized boundary modes may reside either in the bulk gap or fully within the bulk bands of a HOTI, depending on the material details.
Topological insulators that fall into the latter case do not host gapless boundary modes within their bulk bandgap, and as such cannot be distinguished from trivial insulators by their spectrum alone, even with fully open boundary conditions.
This fundamental issue means that HOTIs could be misidentified when their spectra do not exhibit in-gap modes, and it motivates the search for an experimentally measurable indicator of higher-order topology that is protected by \emph{only} spatial symmetries.
It has previously been established that spatial symmetries protect boundary-localized, quantized fractional charge in TCIs \cite{SuSolitons1979,ZakBerry1989,King-SmithTheory1993,VanderbiltElectric1993,RestaMacroscopic1994,HughesInversion2011,TurnerQuantized2012,QuadTheory,CornerCharge,van2018higher}.
In this paper, we demonstrate that a similar feature in metamaterials, namely, the mode density of the spectral bands, can also be fractionally quantized and can indicate both first-order and higher-order topology in gapped TCIs.
We define mode density as the local density of states integrated over an entire band, which is equivalent to the charge density of a filled band in a electronic insulator.
Unlike charge density, using mode density enables us to study the topology of bands without regard to electronic filling or constraints imposed by charge neutrality. 
Here, we experimentally measure the mode densities of two distinct 2D metamaterial insulators  in arrays of coupled microwave-frequency resonators. These two insulators have topology protected by $C_4$ and $C_3$ rotation symmetry respectively.
Although these insulators have broken chiral symmetry and lack in-gap topological modes, we show that fractional quantities in the bulk mode density indicate that both insulators have first-order and second-order topological features protected by rotation symmetry. 
In 2D, we term the quantity indicating second-order topology as a \emph{fractional corner anomaly} in the bulk mode density.
Furthermore, we then show that when these 2D TIs have a non-trivial fractional corner anomaly, they also host topological modes that can be ``pulled" into the bulk bandgap from the bulk bands to form spectrally isolated, corner-localized modes, even when the initial system had no such in-gap modes.
Our results establish a new criterion for identifying higher-order topology in TCIs both theoretically and experimentally, and show that suitable deformations can extract topological corner modes when our criterion is met, even when the original system had no such midgap modes. The latter is crucially important for applications, most of which rely on topologically protected mid-gap modes.

\section{Results}

\subsection{Review of fractional mode density}
\label{Sec:FCA}
\begin{figure}[t!]
\begin{adjustwidth}{-1in}{-1in}
     \centering
     \includegraphics[width = \linewidth]{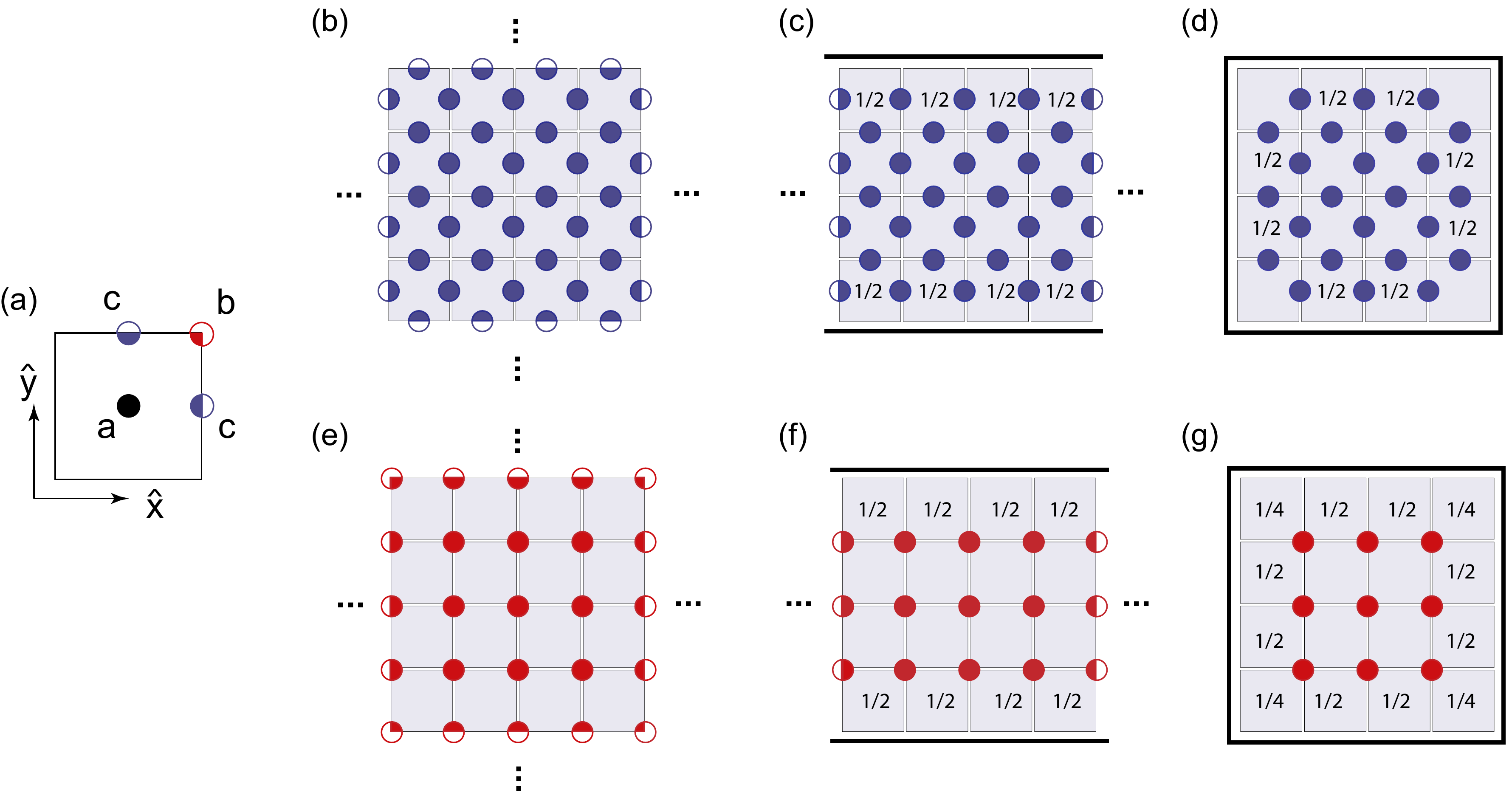}
     \caption{$C_4$ symmetric Wannier center configurations on lattices with different choices of boundary conditions. (a) Wannier centers (represented by circles) at Wyckoff positions that satisfy $C_4$ symmetry. Filled areas indicate the fractional portion of the associated mode density within the unit cell. (b-d) Wannier centers in Wyckoff position $c$ and the fractional portion of mode densities (indicated by numbers) at boundaries per unit cell for a lattice that is: (b) infinite, (c) terminated along $\hat{y}$, (d) terminated along both $\hat{x}$ and $\hat{y}$. (e-g) Wannier centers in Wyckoff position $b$ and the fractional portion of mode densities (indicated by numbers) at boundaries per unit cell for a lattice that is: (e) infinite, (f) terminated along $\hat{y}$, (g) terminated along both $\hat{x}$ and $\hat{y}$.}
     \label{fig_Wannier}
\end{adjustwidth}
\end{figure}

In order to motivate the definition of the fractional corner anomaly, let us briefly discuss the localized Wannier function representation of crystalline insulators, from which we can clearly see how fractional mode densities protected by crystalline symmetry can arise at boundaries.
Wannier functions are a real-space representation of the modes that make up an energy band \cite{WannierStructure1937,MarzariMaximally1997,SouzaMaximally2001}, and they provide a natural basis for studying localized modes in insulators.
TCIs that can be represented by symmetric, maximally-localized Wannier functions are called atomic insulators \cite{bradlyn2017topological,PoSymmetry2017}, and in these materials most of the useful information about the spatial distribution of mode density can be gleaned from just the \emph{Wannier centers}; the center-points of the maximally localized Wannier functions. 
We can further simplify our illustrative discussion by requiring translation symmetry within the bulk, and considering only the zero-correlation-length limit, the latter of which implies that the mode density of each unit cell is simply equal to the number (or fraction) of Wannier centers within it.
We will later show that the concept of a fractional corner anomaly remains well defined outside of this simple limit, and is robust against broken translation symmetry as long as the protective rotation symmetry is preserved.
In symmetric atomic insulators, the Wannier centers associated with an isolated band always form a symmetric configuration.
As a representative example, we consider a $C_4$-symmetric square lattice as shown in Fig. \ref{fig_Wannier}.
In the presence of translation symmetry, the Wannier centers in each unit cell are pinned to symmetric positions (called Wyckoff positions), forming a $C_4$ symmetric configuration \cite{bradlyn2017topological,CanoBuilding2018}.
As shown in Fig. \ref{fig_Wannier}(a), there are three elementary configurations that satisfy $C_4$ symmetry: i) one Wannier center at position $a$, the center of a unit cell, ii) one Wannier center at $b$, the corner of a unit cell, and iii) two Wannier centers at positions $c$, the middle of the edges of a unit cell.
For a Wannier center at Wyckoff position $a$, which is conventionally the topologically trivial position, the associated mode density is fully localized within a unit cell (in the zero-correlation length limit).
When a Wannier center lies between multiple unit cells (at Wyckoff position $b$ or $c$), the associated mode is evenly distributed among these adjacent unit cells, contributing a fractionally quantized mode density to each.
Explicitly, the Wannier center at Wyckoff position $b$ is associated with a mode density of $\frac{1}{4}$ in each of the four adjacent unit cells, and those at Wyckoff position $c$ are associated with $\frac{1}{2}$ mode density in each of the two adjacent unit cells.
Figure~\ref{fig_Wannier}(b, e) shows the nontrivial Wannier center configurations (Wyckoff positions $c$ and $b$, respectively) for bands with $C_4$ symmetry and translation symmetry along $\hat{x}$ and $\hat{y}$, corresponding to bands of an infinite lattice.
This translation symmetry guarantees that the full mode density takes an integer value in every unit cell, even though each Wannier center may contribute non-integer values to each cell.

Interestingly, the local mode density in each cell does not have to take an integer value when the lattice has boundaries. 
Consider a lattice termination along $\hat{y}$ that is consistent with the unit cell structure. This creates two edges parallel to $\hat{x}$ and breaks the $\hat{y}$ translation symmetry at these edges. 
In Figure~\ref{fig_Wannier}(c, f) we show the nontrivial Wannier center representations (Wyckoff positions $c$ and $b$, respectively) that respect $C_4$ symmetry as well as translation symmetry within the bulk for a lattice terminated along $\hat{y}$.
For a unit cell in the bulk, the fractional contributions to the mode density still sum to an integer. 
However, for unit cells at the edges, the mode densities of all relevant Wannier centers sum to give a fractional mode density of $\sigma_y = \frac{1}{2}$ for each of the nontrivial Wyckoff positions (subscript $y$ indicates the termination).
We note that $\sigma$ describes only the fractional portion of the mode density, and is defined as the total mode density in an edge unit cell, modulo 1.
By $C_4$ symmetry, terminating the lattice along $\hat{x}$ would have exactly the same effect for edges parallel to $\hat{y}$.

In the zero correlation-length limit, the fractional mode density is contributed purely from Wannier centers that straddle bulk and boundary unit cells.
In principle, one can decorate the edges with any symmetric configurations of Wannier centers, but this will not change the fractional mode density since these additions always add an integer mode density to each edge unit cell.
The fractional part of the mode density is therefore a property of the bulk and should remain quantized, in units of $\frac{1}{n}$ for $C_n$ symmetry, as long as the bulk symmetries are not broken \cite{CornerCharge}.
These \emph{edge-localized} fractional mode densities are thus the manifestation of first-order nontrivial topology in TCIs, and in an analogous electronic material the equivalent surface charges are indicative of a bulk dipole moment \cite{ZakBerry1989, King-SmithTheory1993, VanderbiltElectric1993, RestaMacroscopic1994, HughesInversion2011}. 
Although the Wannier centers located at Wyckoff positions $b$ and $c$ are associated with the same fractional mode density at edges (i.e., the first-order boundaries), they produce different mode densities at corners (i.e., the second-order boundaries).
Figure~\ref{fig_Wannier}(d, g) shows Wannier center representations (corresponding to Wyckoff position $c$ and $b$, respectively) that preserve $C_4$ and translation symmetries in the bulk on a finite lattice terminated simultaneously along both $\hat{x}$ and $\hat{y}$.
The \emph{fractional portion} of the mode density in the corner unit cells, which we call $\rho$, is zero for Wannier centers in Wyckoff position $c$, and furthermore equals the sum of the edge fractional mode densities, $\rho^{(c)} = \sigma_y + \sigma_x \mod{1} = 0$ for the two edges intersecting at each corner.
In contrast, for Wannier centers in Wyckoff position $b,$ the fractional portion of the mode density is \emph{non-zero} in the corner unit cells, and furthermore \emph{is not} equal to the sum of the edge fractional mode densities, $\rho^{(b)} = \frac{1}{4}\neq \sigma_y + \sigma_x \mod{1} = 0$.
%(see Supplement \S\ref{Sec:wyckoffbc} for more detail). 
%
This quantized, symmetry-protected feature is invisible at boundaries with co-dimension one, but manifests at boundaries of co-dimension two (i.e., at the boundary of a boundary), qualifying it as a higher-order topological feature.
%
%{\bf{TLH: This next statement seems really strong. What do you mean by it? KP: TCIs are protected by spatial symmetry, and FCA is the only higher-order feature protected by spatial symmetry - it is intended to be strong}} 
%
Moreover, it is a \emph{definitive} higher-order topological feature, such that 2D TCIs without a fractional corner anomaly \emph{are not} higher-order.
In the next section, we introduce the fractional corner anomaly as a robust measure of this type of higher-order feature in the bulk mode density.

\subsection{Definition of the fractional corner anomaly}
In the above discussion of the differences between Wannier representations in $C_4$ symmetric TCIs, we found that higher-order topology manifests as anomalous fractional mode density at the corners of 2D lattices that are terminated both along $\hat{x}$ and $\hat{y}$.
For general TCIs with only first-order topology, the fractional mode density $\rho$ in a corner unit cell is the sum of the fractional mode densities, $\sigma_1$ and $\sigma_2$, that respectively manifest at the edges that intersect to form the corner, such that $\rho = \sigma_1 + \sigma_2 \mod{1}$.
A quantized \emph{fractional} deviation from this value indicates higher-order topology.
Therefore, we can define a \emph{fractional corner anomaly},
\begin{equation}
\label{eqFCA}
    \phi \equiv \rho - (\sigma_1+\sigma_2) \mod{1},
\end{equation}
to capture second-order topology in 2D TCIs.
In the zero-correlation-length limit used in the above examples, the fractional mode densities $\rho$ and $\sigma$ are strictly localized within the boundary unit cells and are quantized in multiples of $\frac{1}{n}$ by $C_n$ symmetry.
In general cases with finite correlation length the fractional mode density spreads over an exponentially localized region, such that the mode density in the boundary unit cells is no longer fractionally quantized.
While $\rho$ and $\sigma$ can no longer be respectively defined solely from corner and edge unit cells, we can still calculate a quantized fractional corner anomaly as,
\begin{equation}
\label{eqPhiR}
    \phi = \sum_{{\bf r} \, \in \, \text{one sector}} \phi({\bf r}) \mod{1},\quad 
    \phi({\bf r}) = \mu({\bf r}) - \Big( \mu_1({\bf r}) + \mu_2({\bf r}) \Big),
\end{equation}
where $\mu({\bf r})$ is the local mode density at position ${\bf r}$ when the lattice is terminated along both directions, and $\mu_{1,2}({\bf r})$ are the local mode densities when the lattice is terminated along direction 1 (or 2) and is periodic along direction 2 (or 1) such that no corners exist. 
$C_n$ symmetry divides a lattice into $n$ identical sectors, and the above sum of $\phi({\bf r})$ is performed over one of these sectors.
%
%\Big(See Fig.~\ref{fig_mode_density} for an example for $C_4$ symmetric square lattice.\Big) 
%
We will now show that Eq.~\ref{eqPhiR} is equivalent to Eq.~\ref{eqFCA} in the zero-correlation-length limit, and then argue that $\phi$ cannot change as we increase the correlation length.
In the zero-correlation-length limit, the entirety of the fractional mode density is confined in edge and corner unit cells, such that
\begin{align}
&\sum_{{\bf r} \, \in \, \text{one sector}} \mu({\bf r}) \mod{1} = \rho + \sigma_1 (\ell_1-1) + \sigma_2 (\ell_2-1) \mod{1},\\
&\sum_{{\bf r} \, \in \, \text{one sector}} \Big( \mu_1({\bf r}) + \mu_2({\bf r}) \Big) \mod{1}  = \sigma_{1} \ell_{1} + \sigma_2 \ell_2 \mod{1},
\end{align}
where $\ell_{1,2}$ are the respective lengths of the edges (measured in unit cells) in each sector, and $\rho,~\sigma_{1,2}$ are respectively the fractional mode density in each corner and edge unit cell.
From these identities, it is clear that the definition of $\phi$ in Eq.~\ref{eqPhiR} is equivalent to Eq.~\ref{eqFCA} in the zero-correlation-length limit.
As the correlation length increases from this limit, the fractional portion of the mode density in each sector cannot change, since under $C_n$ symmetry any mode density leaving a sector is exactly compensated by mode density entering that sector.
As a result, the sums of $\mu({\bf r})$ and $\mu_{1}({\bf r}) + \mu_{2}({\bf r})$ over an entire sector, modulo 1, are always constant and quantized in units of $\frac{1}{n}$ \cite{CornerCharge}.
This result, i.e., that the fractional mode density in each sector is equal to the fractional mode density in the zero-correlation-length limit, holds for any finite correlation length as long as $C_n$ symmetry remains.
\begin{figure}[t!]
\begin{adjustwidth}{-1in}{-1in}
\centering
\includegraphics[width=\linewidth]{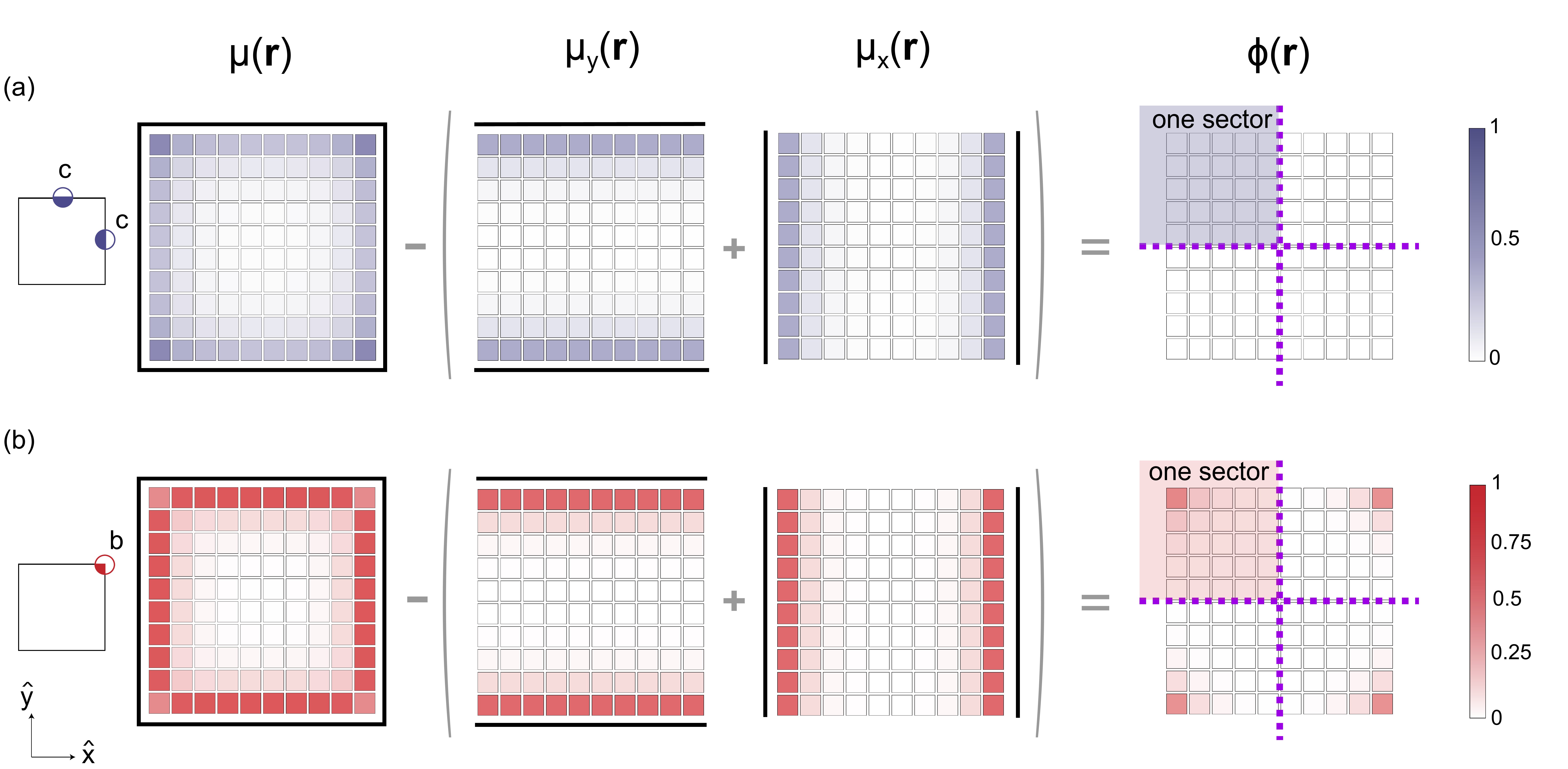}
\caption{
Simulated fractional mode density under various boundary conditions, and calculation of the fractional corner anomaly for (a) A TCI with two Wannier centers at Wyckoff position $c$ showcasing the first-order topology, and (b) A TCI with one Wannier center at the Wyckoff position $b$ showcasing the coexistence of first-order topology and second-order topology. The mode density $\mu({\bf r})$ is calculated for a lattice terminated along both $\hat{x}$ and $\hat{y}$, and the mode density $\mu_{x(y)}({\bf r})$ is calculated for a lattice terminated along $\hat{x}(\hat{y})$. The Hamiltonians of these two TCIs in (a,b) are respectively adapted from generators $h_{2c}^{(4)}$ and $h_{1b}^{(4)}$ in Ref.~\cite{CornerCharge} with symmetric intra-cell coupling terms. }
\label{fig_mode_density}
\end{adjustwidth}
\end{figure}
In Fig.~\ref{fig_mode_density}, we illustrate the calculation of the fractional corner anomaly outside of the zero-correlation length limit for (a) a TCI with Wannier centers at Wyckoff position $c,$ and (b) a TCI with Wannier centers at Wyckoff position $b$.
As stated above, we first calculate $\mu({\bf r})$ on a fully terminated lattice (Fig.~\ref{fig_mode_density}, leftmost column) and $\mu_{1,2}({\bf r})$ on partially terminated lattices (Fig.~\ref{fig_mode_density}, middle columns).
From the partially terminated lattices, we find that fractional mode density manifests at edges for both TCIs, indicating that both have first-order nontrivial topology. 
The rightmost column of Fig.~\ref{fig_mode_density} shows the result of subtracting the mode densities of the partially terminated lattices from the fully terminated lattice, corresponding to the fractional corner anomaly formula in Eq. \ref{eqPhiR}.
The fractional corner anomaly is zero for the TCI in Fig.~\ref{fig_mode_density}(a), indicating that the fractional mode density arises solely from the edges and hence this insulator has \emph{only} first-order nontrivial topology.
For the TCI in Fig.~\ref{fig_mode_density}(b), there is a non-zero fractional corner anomaly, indicating that an additional fractional mode density arises from the corners and hence that this insulator has higher-order topology.
The total anomalous fractional mode density shown in the rightmost column of Fig.~\ref{fig_mode_density}(b) is quantized to $\frac{1}{4}$ per quadrant.
In practice, $\mu({\bf r})$ can always be measured on a fully terminated lattice.
However, it is often not feasible to experimentally create a partially terminated lattice in order to find $\mu_{1,2}({\bf r})$.
In this case, since $\phi({\bf r})$ is localized to corners and decays exponentially, $\mu_{1,2}({\bf r})$ can be approximated as a constant, equal to the average mode density in a narrow section far from any corners and extending from the center of the bulk to the corresponding edge.
Alternatively, if the bandgap is large enough that the correlation length is approximately zero, the fractional corner anomaly can be found experimentally using only the corner and edge unit cells.
Although our discussion above is confined to atomic insulators, i.e., TCIs that admit symmetric Wannier functions, the fractional corner anomaly remains quantized in 2D TCIs that \emph{do not} admit symmetric Wannier functions.
Such TCIs belong to a recently discovered type of ``fragile" topological phase \cite{PoFragile2018,BradlynDisconnected2019,LiuShift2019}.
Fragile TCIs cannot be represented by symmetric Wannier functions, but stacking an auxiliary atomic insulator on a fragile TCI (i.e., coupling the two 2D insulators as if they were stacked) can create an atomic insulator.
Due to this stacking property, the mode density and fractional corner anomaly of a fragile TCI can be found by calculating the difference in mode density between the combined (atomic insulator) system and the auxiliary atomic insulator using our prescription.
Remarkably, although a non-zero fractional corner anomaly does not indicate that corner modes lie within the bulk bandgap, it does indicate the existence of robust topological corner modes somewhere in the spectrum. 
This is because the energy of the topological corner modes is not restricted by the protective crystalline symmetries and is free to take any value, including energies either within the bulk bands (or edge bands) or the bandgap.
When topological corner modes are spectrally isolated from both the bulk and edge modes, they form the familiar exponentially localized 0D in-gap corner modes \cite{Peterson2018, AcousticKagome1, AcousticKagome2, SonicCrystalHOTI, Acoustic3rd1, Acoustic3rd2}.
When not spectrally isolated, the corner modes can generally couple to, and hybridize with, bulk or edge modes, although it was recently shown that corner modes within a bulk band can act as bound states in the continuum (BIC) in the presence of certain symmetries~\cite{benalcazar2019bound}.
Simulation results, detailed in the Supplement \S\ref{Sec:flow}, show that the energy of topological corner modes can be tuned into, and even fully across, the bulk bandgaps (and any edge bandgaps) when a localized potential is applied to only the corner unit cells. 
In the next section, we experimentally measure a non-zero fractional corner anomaly in insulators where the corner modes are hybridized with bulk modes. We then demonstrate that these corner modes can indeed be spectrally isolated and exponentially localized by deformation of the corner unit cell.

\subsection{Experimental Results}
\begin{figure}
	\begin{adjustwidth}{-1in}{-1in}
     \centering
     \includegraphics[width = \linewidth]{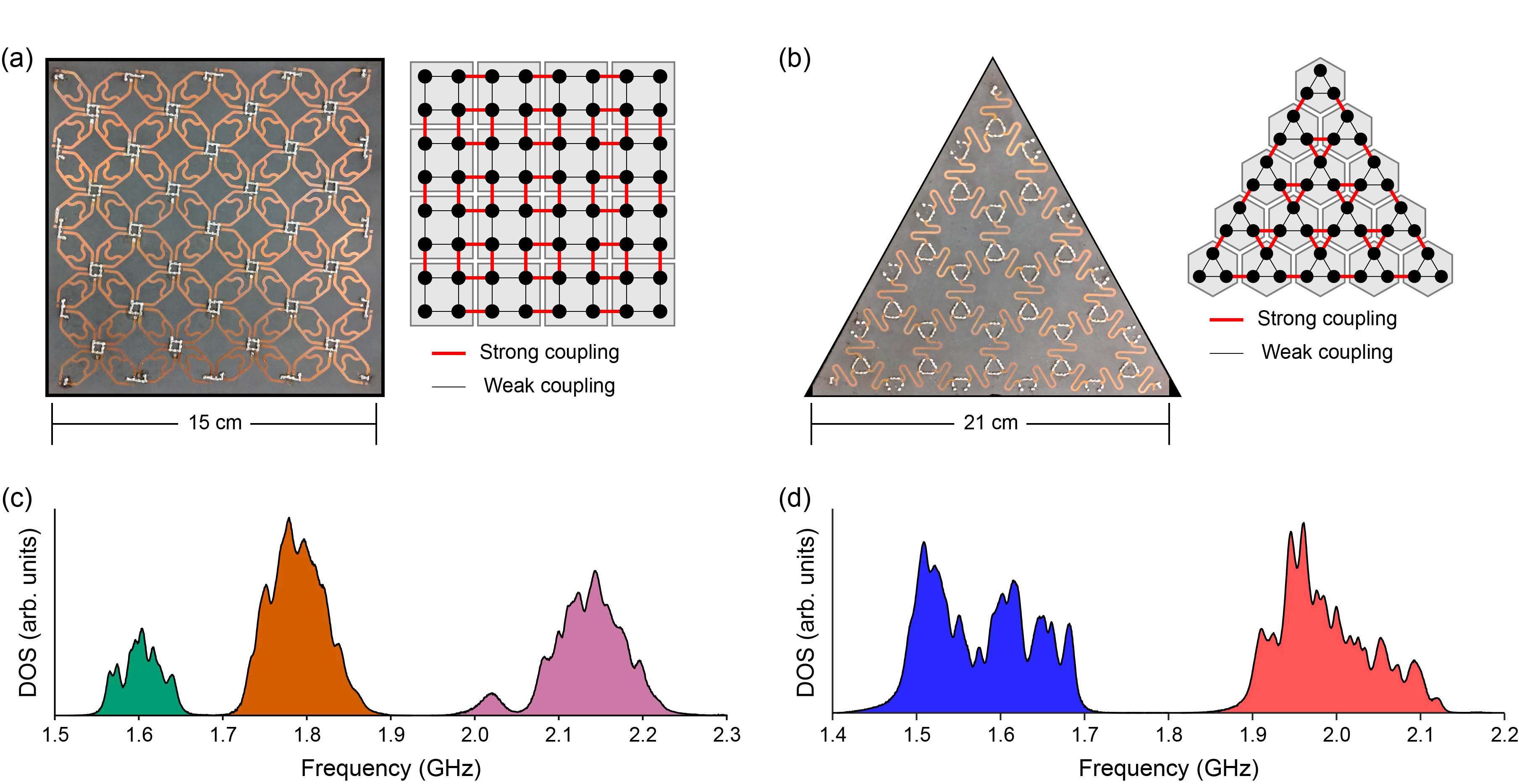}
     \caption{(a) Photograph of the experimental resonator array with $C_4$ symmetry. The schematic on the right illustrates the coupling between resonators. (b) Photograph of the experimental resonator array with $C_3$ symmetry. The schematic on the right illustrates the coupling between resonators. (c) Measured DoS spectrum for the resonator array in (a). (d) Measured DoS spectrum for the resonator array in (b).}
     \label{fig1}
    \end{adjustwidth}
\end{figure}

We experimentally measured the fractional corner anomaly in two rotationally symmetric topological insulator metamaterials implemented using microwave-frequency coupled resonator arrays.
We chose to test two materials with different symmetries because the fractional invariants in TCIs, including the fractional corner anomaly, are related to symmetry representations of the spectrum at high symmetry points of the Brillouin zone. 
We therefore expect to see a different quantization of the fractional mode densities and corner anomalies in these two systems having different rotational symmetry groups.\cite{CornerCharge}
The first insulator, shown in Fig. \ref{fig1}a, is on a square lattice with $C_4$ symmetry, and the second insulator, shown in Fig. \ref{fig1}b, is on a kagome lattice with $C_3$ symmetry.
Both metamaterials consist of half-wavelength microstrip transmission line resonators (similar to those in Ref. \cite{Peterson2018}) with a fundamental resonance frequency of $2.2$~GHz and a typical linewidth of $12$~MHz ($Q \approx 180$).
The coupling between resonators is implemented using discrete capacitors such that the coupling capacitance is $0.1$~pF between resonators in the same unit cell and $0.5$~pF between resonators in neighboring unit cells, which opens the bulk bandgap.
In order to demonstrate that a fractional corner anomaly reflects the fundamental properties protected by only crystalline symmetry, we intentionally did not compensate a coupling-induced difference in the resonance frequencies of resonators at the boundaries compared to those in the bulk.
In essence, the coupling between resonators also causes a shift in their resonance frequency, and at boundaries each resonator is coupled to fewer neighbors than in the bulk.
As discussed in more detail in the Supplement \S\ref{Sec:Chiralbreaking}, this difference in resonance frequencies breaks chiral symmetry (or generalized chiral symmetry for the $C_3$ symmetric models \cite{AcousticKagome2}) and is intrinsically present in many experimental realizations of tight-binding models.
We first found the spectral density of states (DOS) of both metamaterials by means of reflection measurements; see Supplement \S\ref{methods} for details on the measurement technique.
The measured spectrum of the $C_4$-symmetric insulator, shown in Fig. \ref{fig1}c, has three distinct bands.
The measured spectrum of the $C_3$-symmetric insulator, shown in Fig. \ref{fig1}d, has two bands. 
Neither of these insulators have in-gap modes, so from the spectrum alone it is not possible to tell if either metamaterial is topologically non-trivial.
However, as will we show, both are in fact non-trivial, but the intrinsic chiral symmetry breaking causes the edge and corner modes to lie within the bulk bands.

\begin{figure}
	 \begin{adjustwidth}{-1in}{-1in}
     \centering
     \includegraphics[width = 0.9\linewidth]{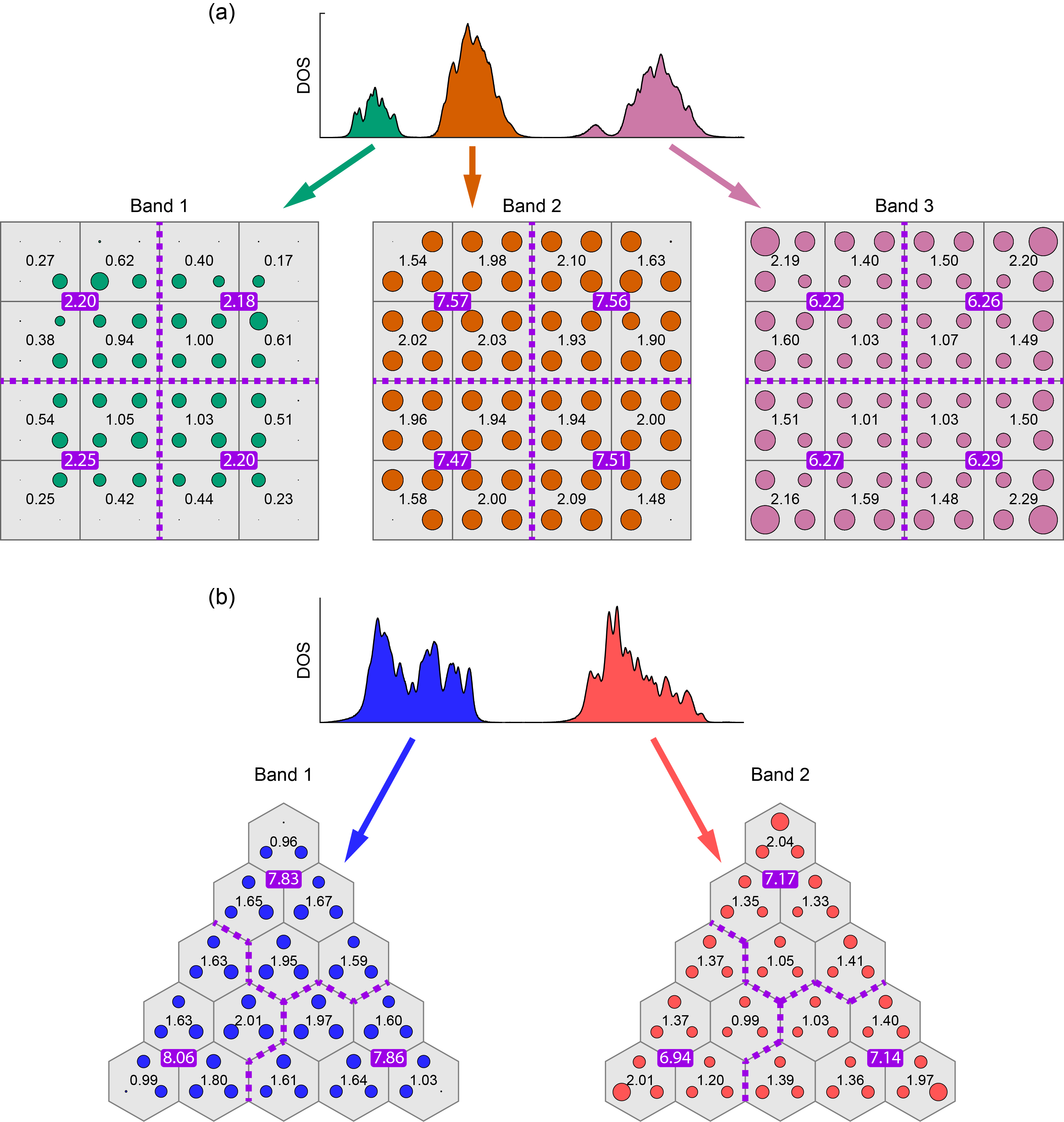}
     \caption{(a) Measured mode density for the $C_4$-symmetric insulator. The mode density for each band is shown separately. Each filled circle represents a resonator, with the area of the circle corresponding to the measured mode density of that resonator in that band. The total mode density of each unit cell is shown with black text, and the total mode density of each sector is shown in purple. (b) Same as (a) but for the $C_3$-symmetric insulator. }
     \label{fig_2}
    \end{adjustwidth}
\end{figure}

We next calculated the mode density of the measured bands by integrating the local density of states in each unit cell over their respective frequency ranges, as shown for both insulators in Fig. \ref{fig_2}. 
The mode density of the $C_4$-symmetric insulators is shown in Fig. \ref{fig_2}a, and has several important features on which we will focus.
First, we find that the resonators in the bulk unit cells are excited in all three bands, indicating that this insulator nominally has three \emph{bulk} bands. 
We observe the total mode density of these bands in each sector is approximately equal, demonstrating that this insulator has approximate $C_4$ rotation symmetry with a small amount of symmetry-breaking disorder.
As expected, the mode density in the bulk unit cells, designated by $\mu^{(4)}$ (where superscript $^{(n)}$ indicates $C_n$ symmetry), is always an integer.
Specifically, $\mu^{(4)}_{1,3} \approx 1$ (where subscript $_m$ indicates band $m$) and $\mu^{(4)}_2 \approx 2$, meaning that band 2 is a two-fold degenerate band while bands 1 and 3 are non-degenerate.
Most importantly, we find a non-zero fractional mode density in the edge and corner unit cells.
Since the band gaps are relatively large in comparison to the width of the bands (i.e., the system is near the zero-correlation-length limit), the entirety of the fractional mode density is tightly localized within the boundary unit cells, and we are approximately in the zero-correlation-length limit.
For bands 1 and 3, the fractional mode density in the edge unit cells is $\sigma^{(4)}_{1,3} \approx \frac{1}{2},$ and in the corner unit cells is $\rho^{(4)}_{1,3} \approx \frac{1}{4}$. In the two-fold degenerate band 2 these fractions are doubled.
This doubling can be understood from the Wannier representation: the two-fold degenerate bulk band has 2 Wannier centers at Wyckoff position $b$ (i.e., band 2 is equivalent to two copies of band 1), and each contributes fractional mode density to the boundary.
Note that the approximate fractions given here are obtained by rounding to the nearest quarter, as we anticipate that $C_4$ symmetry quantizes mode density in fractions of $\frac{1}{n}$ \cite{CornerCharge}.
We can now extract the fractional corner anomaly $\phi$ for each bulk band using the mode density data in Fig.~\ref{fig_2}(a).
Thus, we can find $\phi$ using the simple formula $\phi = \rho - (2 \sigma$), where $\rho$ is the fractional mode density of the corner unit cell, and $\sigma$ is the fractional mode density of the edge unit cells (due to $C_4$ symmetry all edges are expected to be identical).
Here, since there is a small amount of unavoidable disorder in the experiment (which slightly breaks $C_4$ symmetry), we average over all the edges to find $\sigma$, and over all the corners to find $\rho$, such that $\phi^{(4)}_1$, i.e., the fractional corner anomaly for band 1 in the $C_4$-symmetric metamaterial, is
\begin{equation}
    \phi^{(4)}_1 = \rho^{(4)}_1 - 2 \sigma^{(4)}_1 = 0.23 - 0.98 = 0.25 \approx \frac{1}{4}.
\end{equation}
A similar calculation can be carried out for the other bulk bands, giving \[\phi^{(4)}_2 = 0.56 - 0.01 = 0.55 \approx \frac{1}{2}\] for band 2, and \[\phi^{(4)}_3 = 0.21 - 0.02 = 0.19 \approx \frac{1}{4}\] for band 3.
Note that the sum of the fractional corner anomaly over all the bulk bands is always an integer (here rounding gives $\phi^{(4)}_1 + \phi^{(4)}_2 + \phi^{(4)}_3 = 0.99$), and $\phi^{(4)}_2$ is twice $\phi^{(4)}_1$ because band 2 is essentially two degenerate copies of band 1.
These experimentally measured, non-zero fractional corner anomalies indicate the existence of higher-order topology and, as we will show, also indicate the existence of topological corner modes.
In the Supplement \S\ref{Sec:Triv}, we also experimentally measure the fractional corner anomaly for a trivial insulator and show that $\phi^{(4),\text{triv}} \approx 0$ for all bands.
The mode density of the $C_3$-symmetric system is shown in Fig. \ref{fig_2}b.
For this material, we again find that the bulk unit cells have integer mode density, $\mu^{(3)}_1 \approx 2$ (a two-fold degenerate band) and $\mu^{(3)}_2 \approx 1$.
As in the previous experiment, the total mode density of these bands in each sector is approximately equal.
The edge unit cells have a fractional mode density of $\sigma^{(3)}_1 \approx \frac{2}{3}$ in band 1 and $\sigma^{(3)}_2 \approx \frac{1}{3}$ in band 2.
Note that here the approximate fractions are obtained by rounding to the nearest third, since this system is $C_3$ symmetric \cite{CornerCharge}.
Interestingly, although this material does not have a fractional mode density in the corner unit cells, the fractional corner anomaly is non-zero, \[\phi^{(3)}_1 = 0.70 \approx \frac{2}{3} \text{, and } \phi^{(3)}_2 = 0.30\approx \frac{1}{3}.\]

\begin{figure}
	\begin{adjustwidth}{-1in}{-1in}
     \centering
     \includegraphics[width = \linewidth]{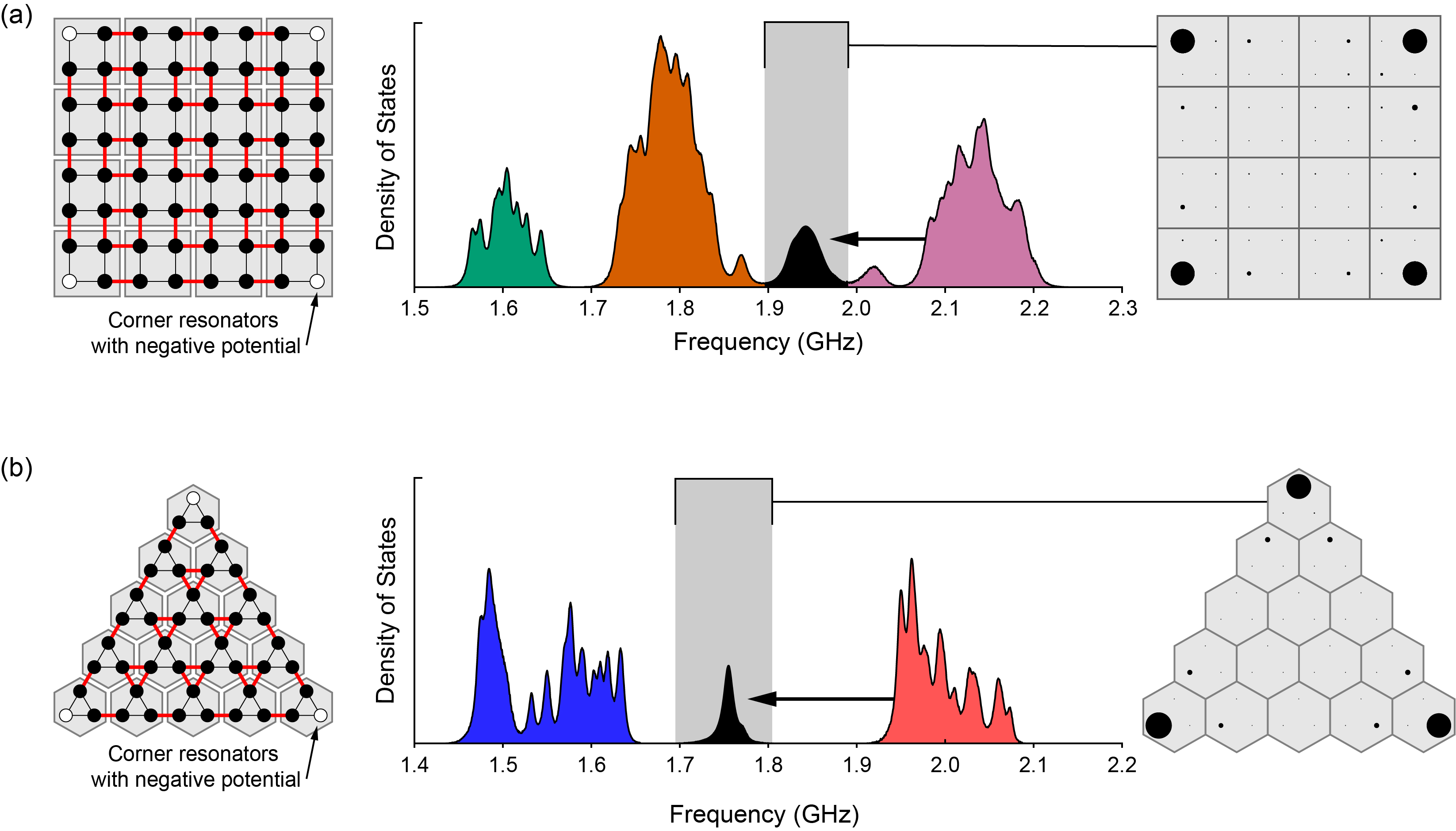}
     \caption{Pulling topological modes into the gap. (a) The schematic on the left shows where a small negative on-site potential is applied (white circles). The measured density of states has modes within the bulk bandgap, which are spatially localized to the corners and confined to one sub-lattice. (b) Same as (a) but for the $C_3$-symmetric insulator.}
     \label{fig3}
    \end{adjustwidth}
\end{figure}

The non-zero fractional corner anomaly in both metamaterials indicates that they are indeed higher-order topological insulators, and we argued above that they should host second-order topological modes at their corners.
Since we have not observed these expected topological modes within the bulk bandgap, we can estimate their approximate energy by finding the band in which the corner resonators are most strongly excited.
In the $C_4$-symmetric system, the corner resonators, around which the second-order topological modes are expected to exist, are mainly excited in band 3, indicating that the corner modes lie in this band. 
Moreover, this implies that we can spectrally localize these modes by slightly lowering the resonance frequency of the corner resonators.
As illustrated in Fig. \ref{fig3}(a), we applied a small negative potential to the corner resonators by means of a capacitor connected to ground, which decreases the electrical length (and thus the resonance frequency) of the corner resonators.
When the potential is applied, the topological modes move into the bandgap between bands 2 and 3 and become exponentially localized to the corner and confined to one sub-lattice.
In the $C_3$-symmetric system, the corner resonators are only excited in band 2, indicating again that the energy of the corner modes is too high and should be lowered to bring the modes into the bandgap.
Again, we pull these modes into the bandgap by similarly applying a small negative potential to the corners, as illustrated in Fig. \ref{fig3}(b).
The topological modes are observed to spectrally localize within the bandgap and spatially localize to the corners with confinement on one sub-lattice.
In the Supplement \S\ref{Sec:Triv}, we conduct a similar experiment on a trivial insulator and show that the modes at the corners of a trivial insulator cannot be spectrally isolated within the bulk bandgap when the same small negative potential is applied to the corners.

\section{Conclusions}

In this work we have argued that in-gap, zero-dimensional boundary modes are not a definitive indicator of non-trivial crystalline topology.
Moreover, we have established the fractional corner anomaly as a superior criterion for generically identifying higher-order topology in theory or experiment. 
This represents a fundamental shift in the understanding of higher-order topological insulators, as the fractional corner anomaly is the \emph{only} real-space feature of bulk bands that directly indicates higher-order topology in TCIs.
From a practical perspective, a focus on bulk-derived fractional mode density could simplify experimental confirmations of novel topological insulators, which often employ ad hoc supplementary boundary elements (e.g. auxiliary resonators or loading capacitors) to spectrally shift topological modes into the bandgap \cite{Peterson2018, AcousticKagome2, Acoustic3rd2, Noh2018, ImhofTopolectrical2018, HuberQuadrupoleNature2018, Xie2019}.
The definition of the fractional corner anomaly can also be extended beyond 2D to identify $d-$th order topology in fully gapped $d$-dimensional insulators.
For example, the fractional corner anomaly for third-order TCIs in 3D is 
\begin{equation}
    \phi = \sum_{i,j} \left( \delta - \rho_j - \sigma_i \right)\mod 1,
\end{equation}
where $\delta$ is corner-localized fractional mode density, $\rho_j$ is hinge-localized fractional mode density, and $\sigma_i$ is surface-localized fractional mode density.
Since this indicator captures fundamental topological features that are protected by spatial symmetries, we expect that it can assist the experimental identification of materials with higher-order topology, which could otherwise be misidentified by only searching for in-gap corner modes.

\section*{Acknowledgements}

The authors would like to thank Prof. Jennifer T. Bernhard for access to the resources at the UIUC Electromagnetics Laboratory, and thank Qingyi Wang and Wentao Jiang for assisting fabrication and measurement of the microwave circuits. This project was supported by the US National Science Foundation (NSF) Emerging Frontiers in Research and Innovation (EFRI) grant EFMA-1627184. C.W.P. additionally acknowledges support from the NSF Graduate Research Fellowship. G.B. additionally acknowledges support from the US Office of Naval Research (ONR) Director for Research Early Career Grant. T.L, W.A.B. and T.L.H. additionally thank the U.S. National Science Foundation under grant DMR-1351895.

\vspace{12pt}
\section*{Author contributions}

C.W.P. designed and fabricated the microwave circuits, performed the microwave simulations and experimental measurements, and produced the experimental figures. T.L. and W.A.B. guided the topological insulator design and performed the theoretical calculations. T.L.H. and G.B. supervised all aspects of the project. All authors jointly wrote the paper.

\newpage

\newcommand{\beginsupplement}{%
        \setcounter{table}{0}
        \renewcommand{\thetable}{S\arabic{table}}%
        \setcounter{figure}{0}
        \renewcommand{\thefigure}{S\arabic{figure}}%
        \setcounter{equation}{0}
        \renewcommand{\theequation}{S\arabic{equation}}%\
        \setcounter{section}{0}
        \renewcommand{\thesection}{S\arabic{section}}%
}

\beginsupplement

\begin{center}
\Large{\textbf{Supplementary Information: \\ A fractional corner anomaly reveals higher-order topology}} \\
\vspace{12pt}
\vspace{12pt}
\large{{Christopher W Peterson}$^1$,
{Tianhe Li}$^2$,
{Wladimir A. Benalcazar}$^3$,
{Taylor L. Hughes}$^2$,
and {Gaurav Bahl}}$^4$ \\
\vspace{12pt}
{\footnotesize{$^1$Department of Electrical and Computer Engineering, University of Illinois at Urbana-Champaign, Urbana, IL, USA}} \\
{\footnotesize{$^2$Department of Physics and Institute for Condensed Matter Theory, University of Illinois at Urbana-Champaign, Urbana, IL, USA}} \\
{\footnotesize{$^3$Department of Physics, The Pennsylvania State University, University Park, PA, USA}} \\
{\footnotesize{$^4$Department of Mechanical Science and Engineering, University of Illinois at Urbana-Champaign, Urbana, IL, USA}}\\
\end{center}

\section{Measuring density of states in a microwave metamaterial} 
\label{methods}
We experimentally find the local DOS of the microwave metamaterials by first measuring the reflection spectrum $S_{11}(f)$ at each resonator, where $f$ is the frequency.
The reflection measurements are taken using a microwave network analyzer (Keysight E5063A). 
The reflection probe is composed of a 50 $\Omega$ coaxial cable terminated in a $0.1$ pF capacitor, which is contacted to each resonator at an anti-node. 
Due to the low probe capacitance, the measured linewidths are dominated by intrinsic losses in each resonator.
The background reflection contributed by the probe is evaluated away from any modes and removed.
This measurement process is similar to that used in Ref. \cite{Peterson2018}.
The absorptance $A(f)$, which is the ratio of absorbed power to incident power, can be calculated from the reflection as $A(f) = 1 - \left| S_{11}(f) \right| ^2$.
To obtain the density of states $D(f)$ for each resonator, we divide the measured absorptance by the frequency squared, $D(f) = A(f) / f^2$, which accounts for increased coupling to the capacitive probe at higher frequencies.
Finally, we normalize $D(f)$ such that $$\int_{\text{all bands}} D_{\textbf r}(f) = ,1$$ where the integration is over the whole band structure and $D_{\textbf r}(f)$ is the local density of states for one resonator, indexed by $\textbf r$. 
Thus, each resonator contributes an overall mode density of $1$, such that for an $N$ resonator system there are $N$ modes total.
Since each resonator supports single participating mode within the measured frequency range, our normalization maps each mode to a mode density of $1$.

\section{Measurement of mode density for a trivial insulator}
\label{Sec:Triv}
\begin{figure}[ht!]
	\begin{adjustwidth}{-1in}{-1in}
    \centering
    \includegraphics[width = 0.7\linewidth]{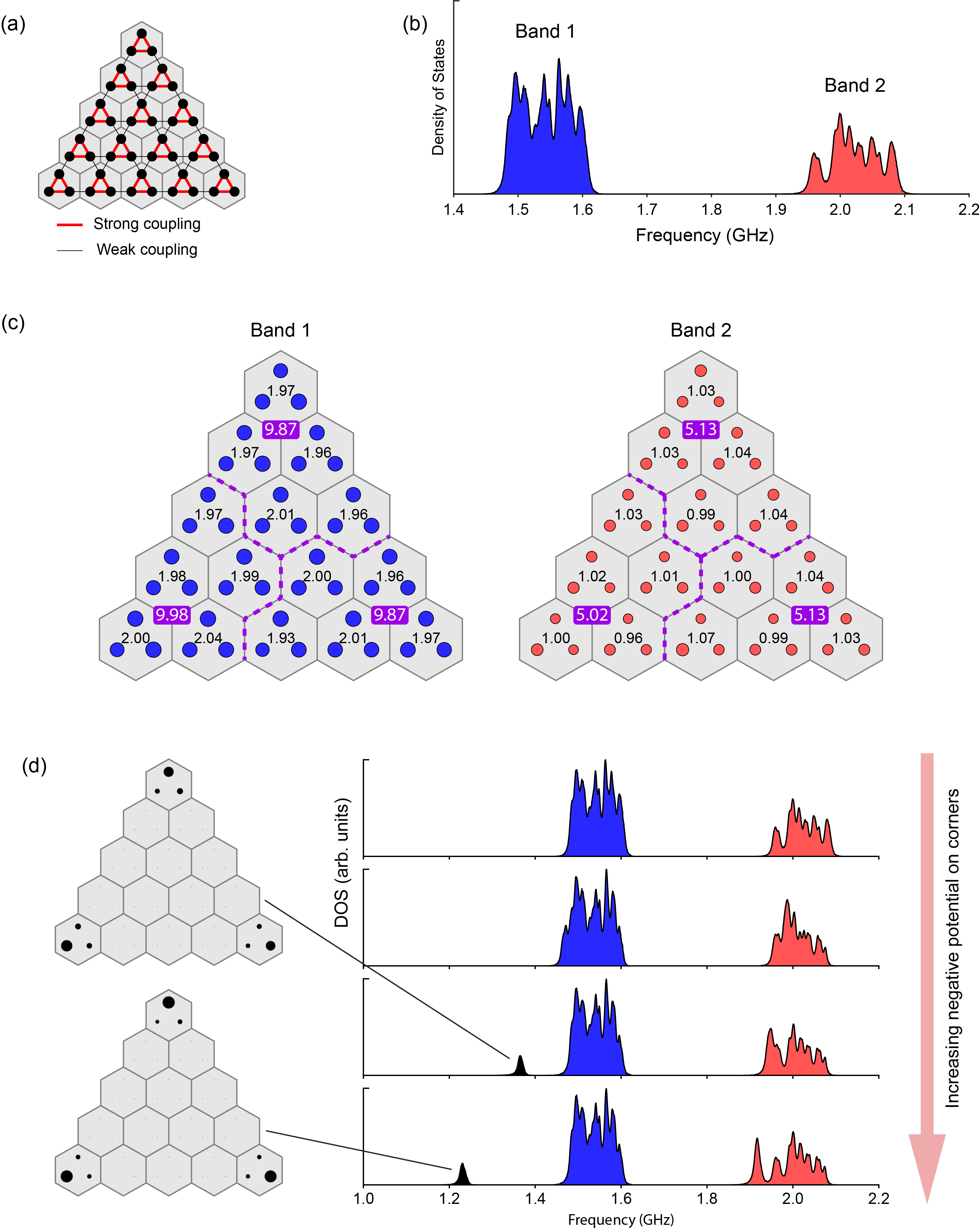}
    \caption{Trivial phase experiment. (a) Schematic of coupled resonator system with strong coupling within unit cells and weak coupling between unit cells. (b) Measured spectral density of states. (c) Measured mode density. Each band is shown separately. Each filled circle represents a resonator, with the area of the circle corresponding to the measured mode density of that resonator in that band. The mode density of each unit cell is shown with black text, and the mode density of each sector is shown in purple. (d) Spectral DoS when a negative potential is applied to the corner resonator. Top row has no negative potential, second row has the same potential as in the main manuscript Fig. \ref{fig3}, bottom two rows have a large negative potential. The measured spatial distributions of the spectrally isolated modes are not confined to the sub-lattice.}
    \label{fig4}
	\end{adjustwidth}
\end{figure}

In this section we experimentally demonstrate that a trivial insulator has 1) integer mode density in all unit cells and 2) no topological corner modes.
We created a trivial $C_3$-symmetric metamaterial insulator on the same board used in the main manuscript by swapping the strong and weak coupling capacitors, such that resonators in the same unit cell are strongly coupled, and those in neighboring unit cells are weakly coupled (Fig. \ref{fig4}(a)).
We again use 0.1 pF capacitors for weak coupling and 0.5 pF capacitors for strong coupling, such that the bandgap is approximately the same size as in the main manuscript.
The measured density of states, shown in Fig. \ref{fig4}(b), is similar to the density of states for the topological $C_3$-symmetric insulator shown in Fig. \ref{fig1}(d), as there are two bulk bands and no observable in-gap states.
In order to fully establish the trivial topology of this metamaterial, we plot the measured mode density for the two bands in Fig. \ref{fig4}(c).
Here, we observe that the mode density takes the same integer value in every unit cell for both bands, $\mu^{(3)}_1 \approx 2$ and $\mu^{(3)}_2 \approx 1$, meaning band 1 is two-fold degenerate and band 2 is non-degenerate.
There are no discernible differences in the mode densities of bulk and boundary unit cells for this metamaterial, clearly indicating that it is topologically trivial.
Since this system is topologically trivial, we expect that there are no topological modes at the boundary of the system. 
To test this, we first applied a small negative on-site potential to the corners of the system (the applied potential is identical to that which was used on the topological insulator in the main manuscript). 
The resulting measured spectrum after this on-site potential is applied is shown in Fig. \ref{fig4}(d).
Unlike in the topological phase, this on-site potential does not pull a localized mode into the bandgap, clearly indicating that there are no topological modes localized at the corners of the system.
We further increased the strength of this on-site potential and found that, with a strong enough potential, localized modes do emerge from the bulk band structure, but these modes do not enter the bandgap.
These modes are not topological modes --- they are tightly coupled to the bulk and require a very strong potential to be pulled out.
Furthermore, the modes pulled out of the band structure are not confined to one sub-lattice, but spread over the corner unit cell as shown in Fig. \ref{fig4}(d).
This is in contrast to a topological mode, which would be confined to a single sub-lattice (as observed in the main manuscript Fig. \ref{fig3}).
Later, in the Supplement \S\ref{Sec:flow}, we discuss pulling modes into the bandgap in more detail.

\section{Spectral flow of high-order topological boundary modes}
\label{Sec:flow}
\begin{figure}[h]
\begin{adjustwidth}{-1in}{-1in}
\centering
\includegraphics[width=\linewidth]{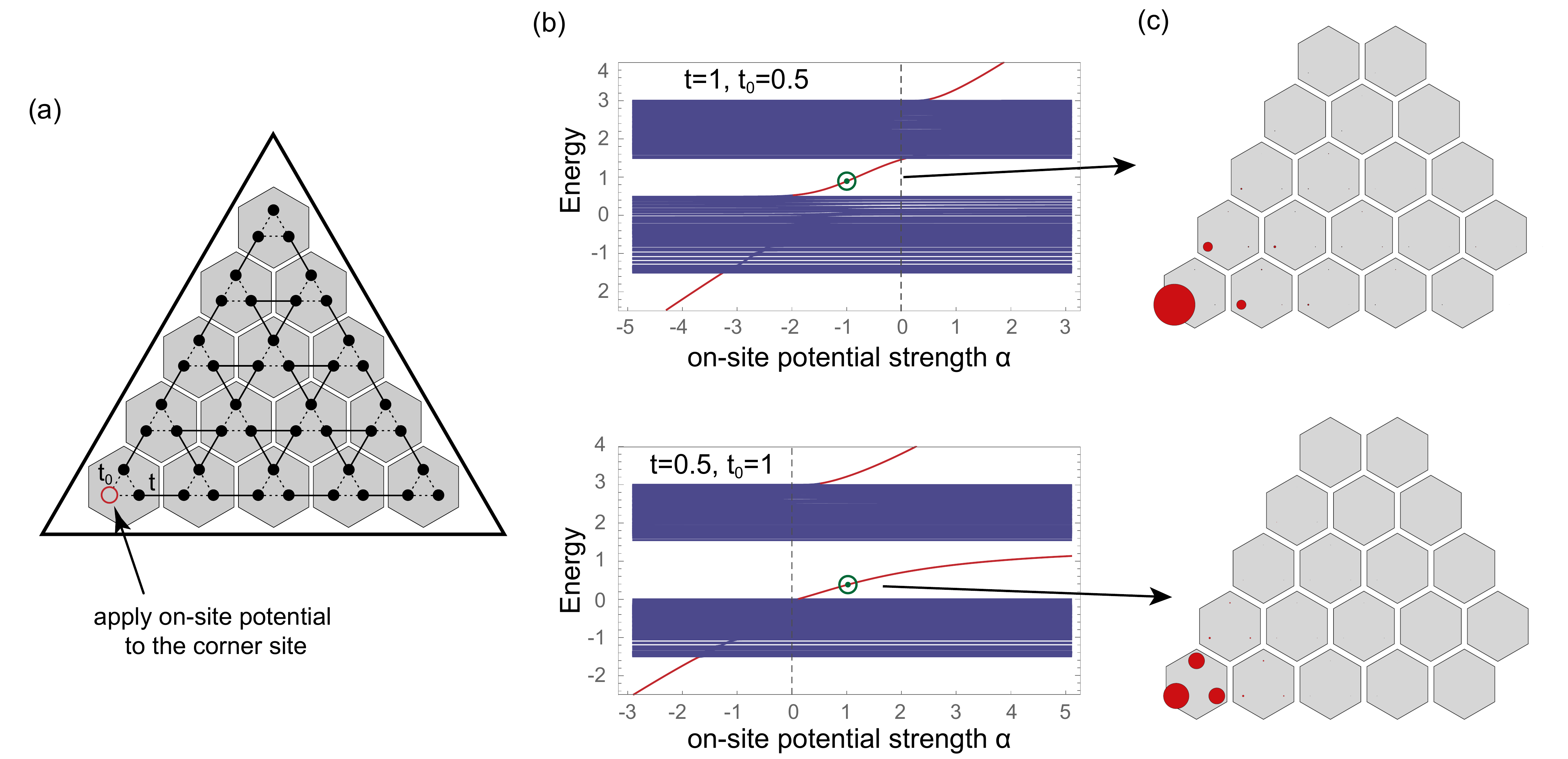}
\caption{(a) Schematic of the lattice configuration for the $C_3$ symmetric TCI. $t_0$ and $t$ indicate the intra-cell coupling strength and inter-cell coupling strength respectively. We apply on-site potential to the lower-left corner site. (b) The spectrum of the $C_3$ TCI as a function of on-site potential $\alpha$ for (top) a topological phase with $t=1,t_0=0.5$ and (bottom) a trivial phase with $t=0.5, t=1$. Red line represents energy shift of the corner mode. (c) The spatial distribution of the in-gap mode (encircled in (b)) over the lower-left corner of the lattice. We use $\alpha=-1$ for the topological phase and $\alpha=1$ for the trivial phase. The simulation is conducted on a triangle lattice with $15$ unit cells per side. }
\label{fig:localization}
\end{adjustwidth}
\end{figure}

In this section, we will take the $C_3$-symmetric TCI that is measured in the main manuscript as an example to illustrate properties of the topological corner modes upon perturbations within corner unit cells. 
The lattice configuration of this TCI is shown in Fig.~\ref{fig:localization}(a). 
When the intra-cell coupling strength $t_0$ is smaller than the inter-cell coupling strength $t$, the TCI is in the topological phase. 
Conversely, it is in the trivial phase when the intra-cell coupling strength $t_0$ is larger than the inter-cell coupling strength $t$. 
Here, we again do not compensate the coupling-induced on-site energy shift on the boundary sites (see \S\ref{Sec:Chiralbreaking} for detailed discussions of this effect), such that in both phases this TCI has two bulk bands and no in-gap modes exist in the spectrum.

We now apply an on-site potential $\alpha$ to the corner site in the lower-left corner unit cell, the red circle in Fig.~\ref{fig:localization} (a), to probe any topological corner modes that may be hidden in the bulk band. 
In Fig.~\ref{fig:localization}(b), we show the simulated spectrum as a function of the on-site potential $\alpha$ for both the topological and trivial phases.
In the topological phase, we observe that one mode (localized at the corner) passes across the entire band structure as $\alpha$ is tuned.
This transfer of a mode between bulk bands, which resembles a topological pumping process, is a unique feature of non-trivial topology.
In contrast, in the trivial phase we observe an avoided crossing of the corner-localized mode with the band structure.
We can pull a mode from the lower bulk band below the band structure when a large negative $\alpha$ is applied, or pull a mode from the upper bulk band above the band structure with a large positive $\alpha$.
However, the mode cannot be transferred across the entire bulk bandgap from one band to the other.
As a further comparison, in Fig.~\ref{fig:localization}(c) we show the simulated spatial distribution of the emergent in-gap modes for both the topological phase and the trivial phase.
In both cases, the in-gap mode is localized at the corner, but this localization alone does not qualify these modes as topological corner modes.
For the trivial phase, the in-gap mode is confined in the corner unit cell and distributes among three sub-lattice sites, indicating a trivial case where the whole corner unit cell is de-tuned from the bulk.
In the topological phase, the in-gap mode is confined to the same sub-lattice site in bulk and edge unit cells, indicating a topological corner mode that is supported by the bulk.

\section{Intrinsic chiral-symmetry-breaking}
\label{Sec:Chiralbreaking}
Chiral symmetry, also known as sub-lattice symmetry, implies that the Hamiltonian $H$ of a system can be broken into two groups, $A$ and $B$, such that coupling terms only exist between them but not within each group.
A chiral symmetric Hamiltonian can be written as,
\begin{equation}
    H = \begin{pmatrix} 0 & H_{AB} \\ H_{AB}^{\dagger}& 0 \end{pmatrix},
\end{equation}
where $H_{AB}$ describes the coupling between the two groups.
In the introduction, we mentioned that chiral symmetry does not protect higher-order topology, but, since it implies that the spectrum is symmetric, it can pin localized modes to zero energy and simplify their observation and the subsequent classification of the material.
For example, if the bulk is gapped at zero energy, chiral symmetry spectrally isolates the corner modes from the bulk such that they are strongly localized to the corners.
Here, we describe how chiral symmetry is intrinsically broken in many metamaterials (as well as natural materials).
This intrinsic symmetry breaking necessitates the mode density approach to classification that we propose in this paper for TCIs in class AI.
Chiral symmetry can appear, at first glance, to be satisfied in any bipartite lattice with nearest-neighbor coupling.
However, a more careful consideration of the tight-binding model shows that, even in these systems, chiral symmetry is intrinsically broken at boundaries.
In tight-binding models, the Hamiltonian at resonator $i$ (or atom, meta-atom, etc.) can be written as 
\begin{equation}
    \hat{H}_i = E^{0}_i + \Delta E_i + \sum_{j \neq i} \gamma_{ij},
\end{equation}
where $ E^{0}_i$ is the intrinsic energy of an isolated resonator, $\Delta E_i$ is the on-site energy shift due to coupling to other resonators, and $\sum_{j \neq i} \gamma_{ij}$ is a sum over the coupling rates between resonator $i$ and all other resonators.
In the microstrip resonators used in our experiments, the on-site energy shift ($\Delta E_i$) of each resonator is equal to the sum of the coupling rates between it and all other resonators, such that $\Delta E_i = \sum_{j \neq i} \gamma_{ij}$.
With periodic boundary conditions and translation symmetry, the first two terms in the Hamiltonian are typically taken as a constant and ignored.
However, since the on-site shift $\Delta E$ is due to the presence of coupled resonators, it is necessarily different for resonators at a boundary, where the surroundings are different (i.e., resonators at a boundary are coupled to their neighbors differently than resonators in the bulk) and where the topological modes of interest are localized.
This difference introduces diagonal terms to the Hamiltonian that cannot be removed, and therefore breaks chiral symmetry and makes the spectrum asymmetric.
In systems with dimerized coupling, where the difference between the strong and weak couplings sets the size of the bulk bandgap, this $\Delta E_i$ shift in on-site energy can never be neglected since it is proportional to the size of the bandgap.
In the classical, designed systems that currently dominate the experimental demonstration of HOTIs, the boundary on-site energy shift is typically compensated in order to regain chiral symmetry.
This is usually accomplished by adding additional elements that shift the energy of the boundary atoms to that of those in the bulk \cite{Peterson2018, AcousticKagome2, Acoustic3rd2}.
However, this approach is problematic for several reasons.
First, the compensation must typically be fine-tuned to match the on-site energy shift of the bulk resonators.
Additionally, if the boundary of the topological phase changes, as in the deformation experiment in Ref. \cite{Peterson2018}, compensation must be added to the new boundary.
Unfortunately, both of these solutions spoil useful features of topological modes, namely that these modes do not require fine-tuning of any parameters and are robust to deformations.
We hope that this work helps to establish a new way of experimentally studying HOTIs that does not require this practice.

\end{document}